\documentclass[twocolumn,aps,prl,showpacs,superscriptaddress]{revtex4}
\usepackage[dvips]{graphicx}
\usepackage{bm}

\begin{document}
\title{Lattice analogy of area-difference elasticity model for lipid-detergent bilayer vesiculation}

\author{L.V.~Elnikova}
\affiliation{
A.I.~Alikhanov Institute for Theoretical and Experimental Physics, \\
B. Cheremushkinskaya st., 25, 117218 Moscow, Russia}
\date{\today}

\begin{abstract}
The vesiculation process was examined in the lipid-detergent
solution
(dimyristoyl-sn-glycero-phoshatidylcholine/octaethylleneglycol
n-dodecyl ether/water), using small-angle neutron scattering
experiments \cite{1}. When observing vesiculation proceeds from
rod-like micelles to unilamellar vesicles, the transformation is
induced by jump-like temperature increase and a monotonic increase
in detergent concentration. Our numerical estimations of the
vesicle shape parameters (the elasticity coefficients and its
fraction on a macroscopic scale) are based upon the area-difference elasticity model \cite{2, 3}. Thus, we composed the
numerical Monte Carlo method, which connects the macroscopic and
microscopic scales by the concept of self-avoiding random
surfaces.
\end{abstract}

\pacs{02.70.Uu, 05.50.+q, 61.30.St, 61.30.Hn, 61.30.Cz}

\maketitle

\section{Introduction}
We study the lyotropic processes, which are widely distributed in
biological systems and define the self-organization phenomena. In
lyotropic liquid crystals, the numerous examples (or even, in its
clear majority), when topological genus of mesophase surfaces was
changed at the phase transitions, are known.

As Kiselev observed in small-angle neutron scattering
investigations \cite{1}, self-assembled vesiculation occurs
between rod-like micelles and unilamellar vesicles with closed
surfaces. Self-assembled vesiculation is induced by temperature
jump of 100 $K\cdot min^{-1}$ from room temperature and increases
as function of detergent concentration. The typical
bilayer-vesicle shape
dimyristoyl-sn-glycero-phoshatidylcholine/water ($DMPC/H_2O$) was
observed experimentally in \cite{5}. The different types of
mesomorphism are distinguished by the nonionic surfactant
influence in
dimyristoyl-sn-glycero-phoshatidylcholine/octaethyllene-glycol
n-dodecyl ether/water ($DMPC/C_{12}E_8/H_2O$) and the drastic
temperature jump.

In principle, certain closed stationary shapes (pears, prolates,
oblates, stomatocytes and nonaxisymmetric shapes with the symmetry
$D_{2h}$) are possible \cite{2}. Theoretical models, which combine
phenomenological and microcanonical methods were developed to
interpret the spontaneous vesiculation phenomena in mesomorphism.
Based on Helfrich theory, a spontaneous curvature $c_0$ (shape
independent $c_0$ with discontinuous budding transition) model and
area difference model (shape independent $\triangle A$ with
continuous one). The area-difference elasticity (ADE) model
\cite{3} describes the energy as a sum of two terms: one depends
on curvature at any point on the surface, and second depends on
curvature-induced area difference between the inner and outer
monolayers. In the bilayer-couple model by Svetina et al.
\cite{6}, the bilayer structure is modelled by representing its
two monolayers as closed neutral surfaces with a constant
separation distance. The elastic energy is composed of the local
and nonlocal bending energies of the membrane.

\section{Model of shape transformations}
At a spasmodic rise in temperature, a rod-like micelle becomes
unstable and transforms into a bilayer, which then collapses into
a hollow sphere. Disintegration of the rod-like micelles and the
bilayer were not experimentally observed \cite{1}, presumably, due
to a fast jump. All of these phenomena look like a first order
phase transitions, and there are situation when the symmetries of
both phases are not in some ratio.

The typical budding process has been phenomenologically described
\cite{7} and the shape equation for the axisymmetric equilibrium
shape was obtained from some general variational ansatz. In a pure
$DMPC$ solution, generative only the lamellar phase, it is found
the lateral compression modulus and bilayer bending rigidity.
However, at addition of nonionic surfactant $C_{12}E_8$, it
becomes difficult to analytically predict the same parameters of
the new mesomorphic series. Using the ADE model \cite{6, 8} it is
possible to determine the minimum energy shape of the vesicle as a
function of the reduced volume $\nu=\frac{V}{V_0}$, where $V$ is
the enclosed volume and $V_0$ is the initial volume. The reduced
volume $\nu$ is concerned with the measured hydrodynamical radius
of a vesicle by certain relations \cite{8}. To define the vesicle
structure parameters, methods that combine the Helfrich
spontaneous curvature theory \cite{3, 8} and the bilayer coupling
model were developed.

In terms of a microscopic canonical ensemble and lattice
Ising-like models, the main mesophases structure parameters may be
connected by a spontaneous curvature, which is constructed on dual
lattices \cite{9} by self-avoiding random surfaces. One would
think so, that in frame of any spin variables, the mixture content
is disregarding.

\section{Lattice Monte Carlo method}
As noted by Caselle \cite{9}, a gas of self-avoiding random
surfaces (SARS) with unconstrained topology and is weighted by the
usual area term $e^{-\beta A}$, where $\textit{A}$ is a surface
area and $\beta=1/k_BT$ ($k_B$ is the Boltzmann constant, $T$ is
the absolute temperature), belongs to the same class of the Ising
model. As other kinds of coupling are introduced, it is possible
to generate these surfaces on the body centered cubic ($bcc$)
lattice and the dual lattice, like the extrinsic curvature. The
reduced Hamiltonian of the Ising model has the form
\begin{equation}
-\beta H=\frac{\beta_h}{2}\sum_{<ij>}\sigma_{ij}
+\frac{\beta_s}{2}\sum_{<kl>}\sigma_{kl}+\frac{\beta_t}{8}\sum_v\sigma_v
\end{equation}
with the partition function
$Z(\beta_s,\beta_h,\beta_t)=\sum_{\sigma=\pm1}e^{-\beta H}$, where
$\beta_s$, $\beta_h$, $\beta_t$ are the coefficients of couplings,
$<ij>$ and $<<kl>>$, $(ijkl)$ are three kinds of coupling, and
$\sigma_i$=-1,0,1 are the spin components, which correspond to the
different types of mixture molecules (lipid, surfactant and water
respectively), $\sigma_v=\sigma_i\sigma_j\sigma_k\sigma_l$. The
coupling coefficients are normalized by action $\textit{A}$ in
terms of the Riemannian curvature $R$ \cite{9}, which is in
accordance to the Gauss-Bonnet theorem $\int
d^2\xi\sqrt{g}R=\sum_vR_v=\chi(S)$. $R_i$ is the curvature
components for four site spin configurations on the $bcc$ lattice.
The concentration term $\beta_t$ in this model is in respect to by
the chemical potential $\mu$ \cite{10}. This treatment allow the
numerical estimations to be conducted with the lattice Monte Carlo
technique for the free energy and structural characteristics of
the vesicular phases.

According to these calculations, all of transitions with
qualitative surface changing genus, at $\beta_t$ range between 0.4 and 0.8,
 may correspond to the appropriate transitions, found in \cite{9} (Fig. 1.). 
The structure parameter of a vesicle shown at Fig. 2.

\begin{figure}
\includegraphics*[width=80mm]{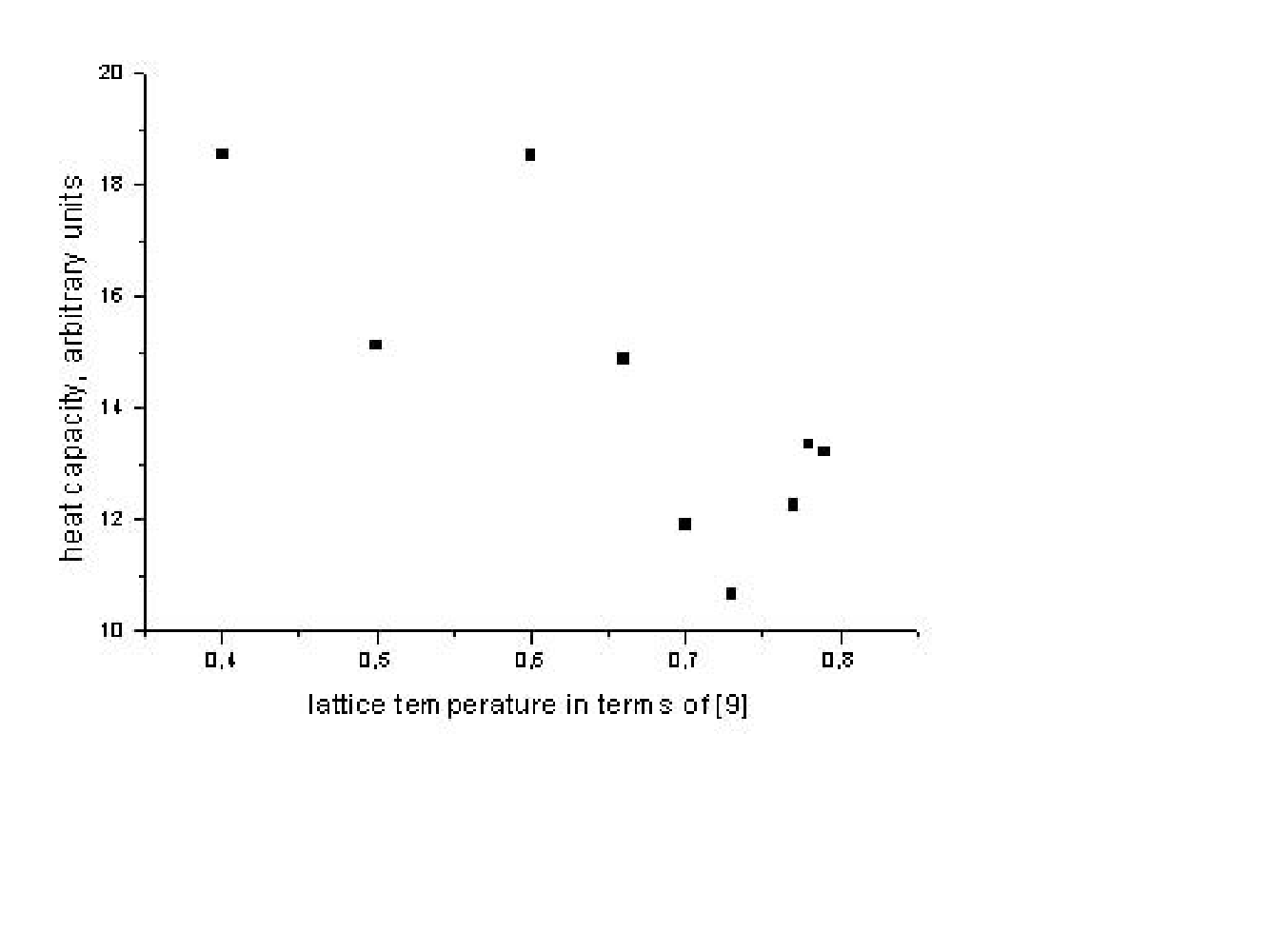}
\caption{\small The heat capacity dependence of temperature
parameter $\beta$ for binary mixture with nonionic surfactant,
Monte Carlo error is $10^{-4}$ }
\end{figure}

\begin{figure}
\includegraphics*[width=80mm]{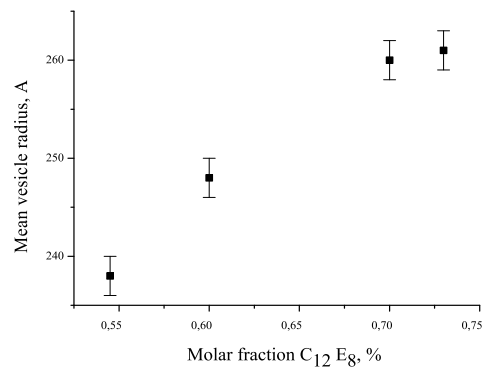}
\caption{\small Monte Carlo simulations for vesicle structural
parameter (mean curvature $R(\AA)$) evolution at 303 K}
\end{figure}

The shape function $r(\theta,\varphi)$ corresponds to the surface
radius $R_s$ for the other follows numerical simulations. This
model conveniently calculates the including volume and other main
structure parameters to plot the phase diagrams of lyotropic
mixtures.

\section{Summary}
The ADE model satisfactorily describes the vesicle phase
transformations, where the non-homogeneous local properties can be
expressed by the elasticity modulus and its fraction in a
macroscopic scale. However, the presented model is well suited for
studying numerous biophysical phenomena and complements the
existing conceptions. In addition to phenomenological ADE
calculations, the lattice treatment of the lyotropic phase
behavior of the aggregates is essential for directly controlling
the concentration, which is a determinative parameter in the
lyomesomorphic transformations. Computer cluster simulation
analysis is useful to investigate the local curvature properties
of different aggregates, including vesicles.

 The author thanks to M.A.~Kiselev for explaning
of the aggregation mechanisms, and also he grateful to
O.P.~Santillan and the anonymous referee for useful correction of
English style.

\end{document}